# Steganalysis: Detecting LSB Steganographic Techniques

Tanmoy Sarkar[*]
Neudesic India Pvt. Limited
Hyderabad, India
tanmoy.sarkar@neudesic.com
* Corresponding Author

Sugata Sanyal
Corporate Technology Office
Tata Consultancy Services,
Mumbai, India
sugata.sanyal@tcs.com

*Abstract*— **Steganalysis means analysis of stego images. Like cryptanalysis, steganalysis is used to detect messages often encrypted using secret key from stego images produced by steganography techniques. Recently lots of new and improved steganography techniques are developed and proposed by researchers which require robust steganalysis techniques to detect the stego images having minimum false alarm rate.
This paper discusses about the different Steganalysis techniques and help to understand how, where and when this techniques can be used based on different situations.**

*Index Terms*— **Steganography, Steganalysis, Stego key, Stego image and Cryptography**

## I. Introduction

With the recent technology advanced people are sharing more and more information among each other's. Some organizations like medicine, military are sharing data with are highly secretive and important. For secure communication people are using cryptography using secret key so that only authenticate receiver can decrypt the message and authentication of message remains intact. But cryptography raised suspicion among attackers and tries to attack the message to get the secretive messages. So, a novel approach of steganography is practiced which contains an innocent cover message embedded with secret message optionally encrypted so that while transferring minimum suspicion arouse among attackers. But, if this approach is used by criminal organization then it becomes necessary to identify the stego multimedia data and try to get the information embedded into it. So, like cryptanalysis which works on cryptography, steganalysis is an art of dissuading covert communication without affecting the innocent ones. The primary purpose of steganalysis is to detect the covert message and if required, try to find more information like secret message length, steganography technique used etc. The issue in steganography and steganalysis is often modeled by the prisoner's problem [20].In this paper we will discuss on various LSB steganalysis approach to attack LSB steganography techniques and there efficiency.

## II. STEGANOGRAPHY TECHNIQUES

Steganography is a method to embed secret image/message into cover image so that the secret message becomes imperceptible to human eyes. To achieve steganography secret message is embedded into cover image using function F (i) and optional stego key to authenticate the data. Kayarkar et al. [16] discuss about various data hiding techniques and there comparative analysis.

Similar to Steganography, another authentication technique is used known as Digital Watermarking. The basic difference between Steganography and digital watermarking is that in digital watermarking the covert data is related to cover data but in steganography the covert data is not related to cover data.

Steganography is mainly distributed among two approaches: reversible and irreversible [14]. Using reversible technique the receiver can extract both the secret message as well as original cover image but while using irreversible technique the receiver can only extract the secret message from stego image leaving original cover image distorted.

Few irreversible techniques are:

1. Battisti et al [1] approach of data hiding using Fibonacci p-sequence number to reduce stego image distortion than traditional LSB technique.
2. Dey et al [2] [3] [21] proposed an improvement over Fibonacci p-sequence LSB data technique of Battisti et al [1] by decomposing pixel value using two approaches: Prime decomposition and Natural number decomposition technique.
3. Nosrati et al. [4] introduced a method that embeds the secret message using linked-list in RGB 24 bit color image

Some reversible data hiding techniques are:

1. Ni et al. [5] proposes a novel approach of data hiding using histogram shifting of original image
2. Kuo et al. [6] presented a reversible technique that is based on the block division to conceal the data in the image.
3. Tian [7] proposes a reversible data hiding technique using difference expansion.

In steganography the simplest method of data hiding is by using LSB steganography method. However, LSB steganography method is divided into two parts: LSB Substitution and LSB Matching techniques. In LSB substitution technique the LSB of cover image is substituted by hidden secret message. In this technique the hidden message bit match with cover image LSB bit. If it is different than substitute the bit otherwise leave as it is. A number of papers have reported very successful steganalysis of LSB replacement [17]–[18]. However, in this approach the even number is always increased but never decreased. Similarly, the inverse is true for odd value. LSB steganalysis methods exploit this asymmetric behavior of LSB replacement to detect secret message.

Let us consider the cover message $m_o$ is grey scale message where each pixel is denoted by 8 bits. So, mathematically to replace the first pixel $x_i$ of message $m_o$ with the first bit of cover message $m_c$ is as follows:

$$x_{LSB_i} = x_i \bmod 2$$
$$x_{covert} = x - x_{LSB_i} + m_c \bmod 2$$

In LSB matching technique the LSB bits are not simply flipped. Instead, the randomly selected sample value is increased or decreased if its LSB does not match the secret message bit to be embedded. The LSB of the cover pixel value finally equals the next bit of the hidden data. Because of this behavior the LSB matching technique is more powerful and difficult to detect then LSB replacement methods. The LSB matching techniques can be denoted as following mathematical expression:

$$p_s = \begin{cases} p_c + 1, & \text{if } b \neq LSB(p_c) \text{ and } (k > 0 \text{ or } p_c = 0) \\ p_c - 1, & \text{if } b \neq LSB(p_c) \text{ and } (k < 0 \text{ or } p_c = 255) \\ p_c, & \text{if } b \neq LSB(p_c) \end{cases}$$

The advantage of LSB embedding is its simplicity and the difference is not visible to naked eyes. But this technique has also having lot of disadvantages like LSB encoding is extremely sensitive to any kind of filtering or manipulation of the stego-image. An attack on the stego-image is very likely to destroy the message. An attacker can easily remove the message by removing (zeroing) the entire LSB plane with very little change in the perceptual quality of the modified stego-image. From Fig. 1.1 we can see that after embedding secret message into the cover image there is significant change in original image histogram pattern suggesting it is being distorted.

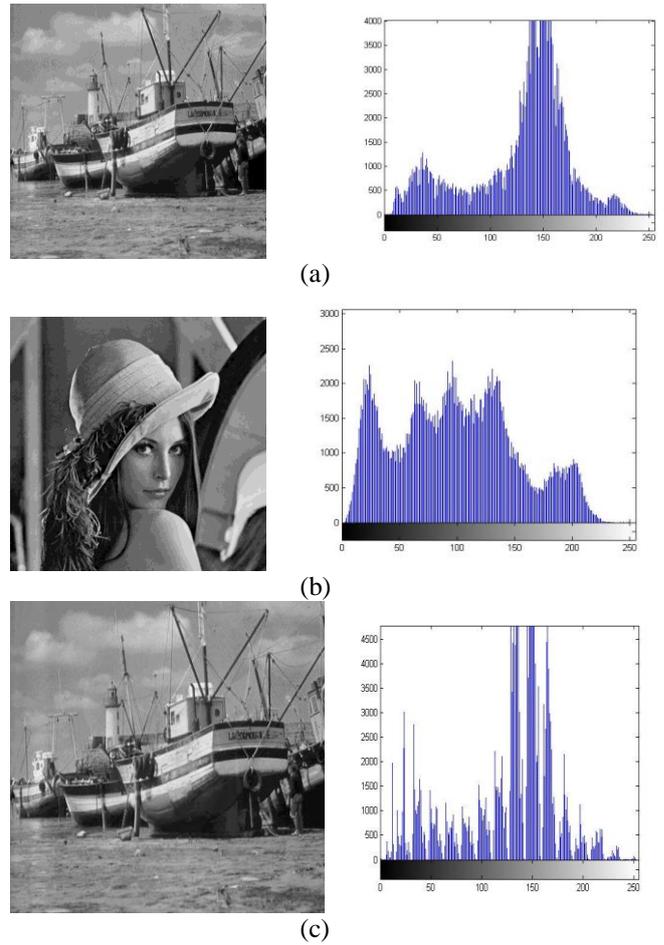

Figure 1.1 LSB Data Hiding Technique (a) Cover Image (b) Image to be embedded (c) Stego Image

### III. STEGANALYSIS TECHNIQUES TO DETECT LSB STEGANOGRAPHY

Steganalysis is an art of detecting the hidden messages inside a stego medium. Steganalysis can be passive or active. In passive steganalysis the attacker tries to detect whether the medium is a stego medium. In active steganalysis the attacker attacks the stego medium, detect and change the stego messages. The presence of embedded messages is often imperceptible but it may disturb the statistics of an image. Determining the difference of some statistical characters between the cover and stego media becomes key issue in steganalysis [19]. There are various steganalysis approaches discuss in paper [15].
Steganalysis algorithm is classified as follows:

1. Specific steganalysis: These types of techniques are established by analyzing the embedding operation and determining certain image statistics. Such techniques need a detailed knowledge of embedding process. These techniques capture very accurate results when used against a target steganography technique. Specific statistical Steganalysis tools are used for finding secret message from stego-images embedded by LSB embedding. The disadvantage of using this method is it is very limited to particular embedding algorithm as well as the image format.

2. Blind / Universal steganalysis: Universal statistical steganalysis comprise the statistical steganalysis method that is not tailored for a specific steganography embedding method. It requires less or even no priori information of the under attack steganographic methods for detection of secret message.

### A. LSB EMBEDDING STEGANALYSIS TECHNIQUES

Fridrich et al. [8] proposes RS algorithm for steganalysis to detect LSB embedding in 24bit color and 8bit grey scale images. In this method the total image pixel is divided into four pixels (2×2 blocks) disjoint groups and applies discriminating function $f(.)$ to these groups to capture the smoothness of these groups.

$f(G)=f(x_1,x_2,……,x_n) = \sum_{i=1}^{n-1}|x_{i+1} - x_i|$ : $x_1,x_2,……,x_n$ are pixels of group G.

Also, three invertible function operations are used on pixel value $x$.
$F_1(x): 0 \leftrightarrow 1, 2 \leftrightarrow 3... 254 \leftrightarrow 255$
$F_{-1}(x): -1 \leftrightarrow 0, 1 \leftrightarrow 2... 255 \leftrightarrow 256$
$F_0(x)$: Identity function

Operation $F_1(x)$ and $F_{-1}(x)$ are applied on group of pixel based on Mask M which is a n-tuple with values -1,0,1. For example, if the values of the four pixels of a group G are 23, 13, 45, 54 and M = (1, 0, 1, 0), then $F_M(G) = (F_1(23), F_0(13), F_1(45), F_0(54))$. The defined function, operation and mask are applied to group of pixels and distributed them into following groups:
    Regular groups: $G \in R \Leftrightarrow f(F(G)) > f(G)$
    Singular groups: $G \in S \Leftrightarrow f(F(G)) < f(G)$
    Unusable groups: $G \in U \Leftrightarrow f(F(G)) = f(G)$

This technique is based on analyzing how the number of regular and singular groups changes with the increased message length embedded in the LSB plane. The greater the message size, the lower the difference between $R_{-M}$ and $S_{-M}$ and the greater the difference between and $R_M$ and $S_M$. This behavior is used in detection of hidden message from the stego-image

Westfeld et al. [9] proposed method based on statistical analysis of Pairs of Values (PoVs) using chi-square attack that is exchanged during message embedding. The idea of statistical analysis is to compare the theoretically expected frequency (which is the arithmetic mean of two frequencies in a PoV since we do not have original cover medium) distribution in steganograms with some sample distribution of stego medium.
To get the statistics for the difference between distributions of PoV's used the below mentioned equation:

$$x_{PoV}^2 = \sum_{i=1}^{127} \frac{\left(Y_{2i} - \frac{1}{2}(Y_{2i} + Y_{2+!})\right)^2}{\frac{1}{2}(Y_{2i} + Y_{2+!})}$$

The probability of embedding is decided by calculating the p-value defined below:

$$p = \Pr(x_{k-1}^2 \geq x_{PoV}^2) = 1 - \frac{1}{2^{\frac{k-1}{2}} \Pi\left(\frac{k-1}{2}\right)} \int_0^{x_{PoV}^2} e^{-\frac{x}{2}} x^{\frac{k-1}{2}} dx$$

This p-value is calculated for a sample from the values examined which starts from the beginning of the image and gets amplified for each measurement.
This method provides very reliable results when the message placement is known (e.g., sequential). However, randomly scattered messages can only be reliably detected with this method when the message length becomes comparable with the number of pixels in the image.

Kerke et al. [10] uses a gray level co-occurrence matrix to identify LSB embedded images. Each element $(i,j)$ in grey level co-occurrence specifies the number of times that the pixel with value $i$ occurred horizontally adjacent to a pixel with value $j$. In this technique four different co-occurrence matrix is generated using four different direction respectively. Then the average of each four matrix is taken. Since the gray level pixel are correlated they are tend to be concentrate on diagonal. As the data embedding increase the pixels are spreads more as the correlation decreases. The author created $2^n$ feature vectors where n is number of diagonals of co-occurrence matrix. Classification of images as stego is determined by using Absolute distance and Euclidean distance by using following steps:
1. Feature vector of the test image is generated
2. Absolute distance measure and Euclidean distance measure is used to check the closeness of the test image and the training database images.
3. Distance values are sorted in ascending order and minimum of the values is considered
4. A threshold value is set to determine whether the image is stego image.

### B. LSB MATCHING STEGANALYSIS TECHNIQUES

Zang et al. [11] proposes a method of LSB Matching steganography by detecting increasing local minima and decreasing local maxima of histogram of stego. In this method the intensity of histogram is defined as:

$$h_c(n) = |\{(i,j)| p_c(i,j) = n\}|$$

: $h_c(n)$ indicates the pixel value of intensity n.

Considering the fact that embedding locations are uniformly distributed and independent of pixel value the embedding rate $\rho$ is applicable to only 50% of LSB value because the rest has already had the desired value. Hence, 1-$\rho$/2 of the pixels have not been modified. The histogram of stego image is given as:

$$h_s(n) = \left(1 - \frac{\rho}{2}\right) h_c(n) + \frac{\rho}{4}\left(h_c(n-1) + h_c(n+1)\right)$$

Based on this equation the following conditions are derived:

$h_s(n^*) < h_c(n^*)$ for local extremum
$h_s(n_*) > h_c(n_*)$ for local minimum

This condition helps in detecting steganography in images.

Harmsen [12] exploit the histogram characteristic function (HCF) to detect of steganography in color images. Ker [13] improved Harmsen's method through: (i) introduce a calibration mechanism by using a down sample technology, and (ii) substitute the usual intensity histogram for the adjacency histogram to compute the histogram characteristic function.

1. Calibrated HCF COM Detector

Consider down sampling an image by a factor of two in both dimensions using a straightforward averaging filter. Precisely, let be the pixel intensities of the down sampled cover image $p'_c(i,j)$ and similarly down sampled version of the stego image $p'_s(i,j)$ is given by:

$$p'_c(i,j) = \left\lfloor \sum_{u=0}^{1} \sum_{v=0}^{1} \frac{p_c(2i+u, 2j+v)}{4} \right\rfloor$$

$$p'_s(i,j) = \left\lfloor \sum_{u=0}^{1} \sum_{v=0}^{1} \frac{p_s(2i+u, 2j+v)}{4} \right\rfloor$$

They divide the summed pixel intensities by four and take the integer part to reach images with the same range of values as the originals. They compute the HCF and COM of these two down sampled images $C(H'_c[k])$, $C(H'_s[k])$ using below equation and they use:

$$C(H[k]) = \frac{\sum_{i=0}^{n} i |H[i]|}{\sum_{i=0}^{n} |H[i]|}$$

Finally they conclude that if the image is not a steganography image then $C(H'_c[k]) \approx C(H_c[k])$ i.e. the Centre of Mass is nearly equal even after sampling the cover medium. The variation between the magnitude of the values they use $C(H_c[k]) / C(H'_c[k])$ as dimensionless discriminator.

2. Compute Adjacency Histogram

Here they consider the two dimensional adjacency histogram and since adjacent pixels are close to intensity the histogram is sparse off the diagonal. The intensity $h_c(n)$ is given by:

$$h_c(n) = |\{(i,j) | p_c(i,j) = n\}|$$

This histogram of sparse value off the diagonal is affected in case of steganography.

## II. CONCLUSION

In this paper we have discussed LSB steganalysis techniques and there line of attack on images. This paper helps to understand LSB steganography and its different steganalysis methods which further help in future studies.